\documentclass[12pt]{article}
\usepackage{epsfig}
\begin{document}
\def\sech{\mathop{\rm sech}\nolimits}
\def\etal{{\it et al.}}
\def\today{\number\day
           \space\ifcase\month\or
             January\or February\or March\or April\or May\or June\or
             July\or August\or September\or October\or November\or December\fi
           \space\number\year}
\def\thisday{21 October 1996}
\begin{titlepage}
\begin{flushright}
Fermilab-Pub-96/358-T\\
hep-ph/9610433\\
\end{flushright}
\vspace{1cm}
\begin{center}
{\Large\bf Next-to-Leading Order Gluonic Three Jet Production 
at Hadron Colliders }\\
\vspace{1cm}
{\large
William B. Kilgore \\
\vspace{0.5cm}
and \\
\vspace{0.5cm}
W.~T.~Giele}\\
\vspace{0.5cm}
{\it
Fermi National Accelerator Laboratory, P.~O.~Box 500,\\
Batavia, IL 60510, U.S.A.} \\
\vspace{0.5cm}
{\large \thisday}
\vspace{0.5cm}
\end{center}
\begin{abstract}
We report the results of a next-to-leading order event generator of purely
gluonic jet production. This calculation is the first step 
in the construction of a full next-to-leading order calculation 
of three jet production at hadron colliders.
Several jet-algorithms commonly used in experiments are implemented and
their numerical stability is investigated.
\end{abstract}
\vfill
\end{titlepage}
\section{Introduction}
In this paper we report the first step in constructing a 
Next-To-Leading order (NLO) three jet event generator for hadron colliders. 
This involves the construction of the pure gluonic contribution to 
this cross section. The calculation combines the
one loop virtual matrix elements $gg\rightarrow ggg$ \cite{BDKa} with the real 
matrix elements $gg\rightarrow gggg$ \cite{PT,Ku,GuKa,BGa,MPX,MP}. 
The major issue we want to address in this letter is the convergence 
and numerical stability of the NLO event generator. The jet algorithm is an
integral part of the observed final state and is needed to define the NLO
3-jet cross section. Unlike the NLO 2-jet calculation, the NLO 3-jet
calculation is sensitive to many details of the jet algorithm.
This is because of the presence of the 4-parton final state, which
by applying the jet algorithm is converted into either a 2-, 3- or 4-jet
final state. A complete understanding of this partitioning into
different numbers of jets requires a careful study of the details of
different jet algorithms. For this paper we consider four algorithms :
\begin{itemize}
\item[(a)] The ``fixed-cone'' algorithm, used by UA2 \cite{UA2alg}
\item[(b)] The ``iterative-cone'' algorithm, used by CDF \cite{CDFalg} 
           and D0 \cite{D0alg}.  
\item[(c)] The ``$K_T$'' algorithm \cite{KtalgTh}, 
           under study by CDF and D0 \cite{KtalgExp}.  
\item[(d)] The ``EKS'' algorithm, used in NLO 1-jet and 2-jet 
           inclusive calculations \cite{EKSalg}.
\end{itemize}

In section 2 we will describe the methods and techniques
used in the event generator in some detail. Section 3 describes 
and investigates the stability of the four jet-algorithms. 
Some distributions are shown in section 4 as an illustration of
the achievable numerical accuracy of the event generator.
No attempt is made for a detailed phenomenological study;
this only makes sense once the quark contributions have been included. 
Finally, in section 5 we summarize the findings of the study.

\section{The method}

The construction of a flexible event generator requires the 
generation of partonic final states with a minimal amount of 
implicit phase space integration. At Leading Order (LO) this
is trivial, but at NLO it requires careful handling 
of the cancellation of divergences between the
soft/collinear contributions and the virtual corrections. 
The divergences stem from the fact that at NLO a parton can only be defined
through a resolution criterion. This resolution criterion can take many forms, 
from a simple invariant mass cut to 
a full blown fragmentation function. For the studies in this paper
a simple invariant mass resolution criterion, $s_{min}$, suffices. 
That is, if the invariant mass of two partons is smaller 
that $s_{min}$ they are considered to be unresolvable and treated 
as a single parton by integrating out the unresolved
phase space. This isolates the unresolved soft/collinear region
of phase space from the resolved bremsstrahlung phase space.
After this rearrangement, both the resolved contribution and the
combination of the unresolved 
soft/collinear contributions with the  virtual corrections
are finite \cite{GG,GGK}.

With the above method it is easy to calculate the soft/collinear contributions.
The next step is to use this calculation to construct a NLO event generator.
There are in principle three methods of putting together the resolved 
partonic cross sections in order
to make the NLO jet event generator. In order of complexity
they are:
\begin{itemize}
\item[(a)] ``The slicing method'', in which both matrix element 
           and phase space are approximated \cite{Owens}
           in the soft/collinear region.
\item[(b)] ``The subtraction method'', in which the phase space is 
           still approximated in the soft/collinear region, 
           but the matrix element is now exact (by adding in 
           the correction factor numerically) \cite{EKSsub}.
\item[(c)] ``The exact method'', in which both the correction factors 
           for the phase space and matrix elements in the unresolved 
           region are added in numerically \cite{CatSey}.
\end{itemize}

Method (a) is used to analytically calculate the soft/collinear region. To be
able to perform the integrations and extend the method to arbitrary partonic
processes one has to approximate both the matrix element and the phase space
in the soft/collinear region.
For any useful and numerically stable event generator method (b) is often
sufficient. In a numerical calculation it is
trivial to extend method (a) to method (b).
Method (c) is attractive because
there are no approximations. That is, no terms of order $s_{min}$ 
have been neglected and one can choose the resolution parameter 
as large as one wants without changing the results.
This method, however, is more cumbersome to implement.

One can describe the different methods better using a schematic formula. 
The $n$-parton contribution to the $(n-1)$-jet cross section is
given by
\begin{eqnarray}
\label{eq:dsign}
d\,\sigma_n &=& |{\cal M}_n|^2\times J_n\ d\,\mbox{PS}_n \nonumber \\
            &=& \left(|{\cal M}_n|^2\times (1-\theta_s)
               +|{\cal M}_n|^2\times\theta_s)\right)
               \times J_n\ d\,\mbox{PS}_n \nonumber \\
	    &=& |{\cal M}_n|^2\times (1-\theta_s)J_n\ d\,\mbox{PS}_n
               +\theta_s\times
               \left(T_1(\theta_s)+T_2(\theta_s)+T_3(\theta_s)\right)\ ,
\end{eqnarray}
where the $n$-parton differential cross section $d\,\sigma_n$ is given by
the matrix element squared, $|{\cal M}_n|^2$, and the phase space
constraints from the jet algorithm and cuts, $J_n$, integrated over
the $n$-parton phase space $d\,\mbox{PS}_n$. The soft/collinear
unresolved part of phase space is separated off using the resolution
criterion embodied in the quantity $\theta_s$, which takes the value
$\theta_s=1$ in the unresolved phase space region and  $\theta_s=0$
otherwise.

$T_1$ is given by
\begin{eqnarray}
\label{eq:T1}
T_1(\theta_s) &=& 
S\,|{\cal M}_{n-1}|^2\times J_{n-1}\ d\,\mbox{PS}_{soft}\ d\,\mbox{PS}_{n-1}
\nonumber \\
    &=& R(\theta_s)\,|{\cal M}_{n-1}|^2\times J_{n-1}\ d\,\mbox{PS}_{n-1}\ ,
\end{eqnarray}
and represents the integral of the approximate matrix element $|{\cal
M}_n|^2 \rightarrow S\, |{\cal M}_{n-1}|^2$ over the approximate phase
space $d\,\mbox{PS}_n\rightarrow d\,\mbox{PS}_{soft}\
d\,\mbox{PS}_{n-1}$.  The resolution factor $R(\theta_s)$ is
independent of the hard scattering and can be calculated analytically
for a wide range of multiparton processes \cite{GG,GGK}.  $T_2$ is
given by
\begin{equation}
\label{eq:T2}
T_2(\theta_s) = \left( |{\cal M}_n|^2 - S\,|{\cal M}_{n-1}|^2\right)
\times J_n\ d\,\mbox{PS}_n\ ,
\end{equation}
and represents the integral over the true unresolved phase space of the
difference between the true matrix element and the approximate matrix
element. $T_3$ is given by
\begin{equation}
\label{eq:T3}
T_3(\theta_s) = S\,|{\cal M}_{n-1}|^2\left( J_n\ d\,\mbox{PS}_n
      -J_{n-1}\ d\,\mbox{PS}_{n-1}\,d\,\mbox{PS}_{soft}\right)\ ,
\end{equation}
and represents the difference between the integrals of the approximate
matrix element over the true unresolved phase space and the
approximate unresolved phase space. Note that $T_1$ contains the soft
and collinear divergences needed to cancel the singularities of the
virtual term, while $T_2$ and $T_3$ vanish as the domain of support
for $\theta_s$ is taken to zero.

Method (a) keeps $T_1$, but sets $T_2 = T_3 = 0$,  method (b) keeps
both $T_1$ and $T_2$, but sets $T_3 = 0$, while method (c) keeps all
three terms.
The terms proportional to the soft factor $S$ 
cancel between $T_2$ and $T_3$
so that the final expression for method (c) is somewhat simplified. 
The advantage of method (c) is that the $\theta_s$-dependence exactly
cancels for any value of this resolution parameter.  
The drawback is that apart from the usual negative 
weighted virtual plus soft/collinear and 
positive weighted bremsstrahlung contributions we have now an 
additional type of negative weighted events which numerically 
cancel the subtraction term $R(\theta_s)$. 
This can often be confusing, especially when one chooses large values
of $\theta_s$, because one has a different phase space constraint on
this type of bremsstrahlung term. Using method (b) removes these  
additional events, but now we must choose $\theta_s$ to be
sufficiently small that the phase space approximations are valid. In
general this poses no problem and in practice 
this is the method we use. The effects of the three methods can 
easily be demonstrated numerically. The $s_{min}$-dependence
of methods (a) and (b) are shown in fig.~\ref{fig:smin} for several
jet algorithms. We postpone the discussion of these dependences to
section 4.

\section{Jet Algorithms}

The purpose of the jet algorithm is to 
quantify certain topological features of
hadronic energy flow in scattering processes. 
By identifying high transverse
momentum hadronic clusters in collisions
we can make a connection with the
underlying partonic scattering and apply 
perturbative QCD to predict the cross section. 
The form of the jet algorithm depends to a 
large extent on the capability
of the detector and on the collision environment. 
Theoretical issues are only of secondary importance. 
A stable experimental jet algorithm is, by definition, theoretically
infrared safe. There are of course issues of perturbative convergence,
but the experiment (and implicitly the data) 
should determine the jet algorithm not vice versa.

With current techniques for theoretical calculations one can easily 
accommodate any stable experimental jet algorithm. 
The only crucial theoretical issue is a reliable estimation of the 
theoretical uncertainties. This is why the NLO predictions for 
observables are so important.
By comparing NLO with LO we can determine the regions of 
phase space where we can make
reliable predictions and give estimates of the uncertainty. 
There is no point ``improving'' predictions without 
a clear understanding of the theoretical 
uncertainties in the ``improved'' predictions.

The extension of the NLO 2-jet calculation to NLO 3-jet is
non-trivial with respect to the jet algorithm
as we will now explain. 
The algorithms usually depend on a cone-size or distance scale between
the clusters:
\begin{equation}
\label{eq:Rdef}
R = \sqrt{(\Delta \eta)^2 + (\Delta \phi)^2}\ ,
\end{equation}
where $\Delta\eta$ is the difference in pseudorapidity and
$\Delta\phi$ the difference in azimuthal angle.
When combining clusters of energy one usually uses transverse
energy-weighted ($E_T$-weighted) clustering:
\begin{eqnarray}
\label{eq:ETmerge}
E_T^{TOT}&=& \sum_i\, E_T^{(i)} \nonumber\\
<\eta>&=&\frac{1}{E_T^{TOT}}\sum_i\, E_T^{(i)}\eta_i \\
<\phi>&=&\frac{1}{E_T^{TOT}}\sum_i\, E_T^{(i)}\phi_i\ . \nonumber
\end{eqnarray}
We will now summarize our implementations of the four jet algorithms
under consideration:
\begin{itemize}
\item[(a)] \underline{The ``fixed-cone'' algorithm} \newline
This algorithm was used by UA2 and is 
described in some detail in ref. \cite{UA2alg}.
This algorithm
is the most basic and straightforward of all the 
four algorithms we are considering.
The procedure is very simple:
\begin{itemize}
\item[1.] Form a cluster list, ordering all clusters by $E_T$.
\item[2.] Select the highest $E_T$-cluster from the cluster-list and
          draw a cone of radius R around the cluster axis.  Calculate
          the transverse jet energy and a new jet-axis by performing
          the $E_T$-weighted sum of all the clusters in the cone as
          defined in eq.~\ref{eq:ETmerge}.
\item[3.] Remove all clusters in the cone from 
          the cluster-list and move the jet to the jet-list.
\item[4.] If the cluster-list is not empty go to step 2.
\item[5.] Apply the appropriate minimum transverse energy and rapidity cuts 
          to the entries in the jet-list to find the final set of
          jets.
\end{itemize}
Note that all of the basic physics involved in the clustering is
already contained in the 3 parton final states (i.e. NLO 2-jet
production or LO 3-jet production).  No matter how many additional
partons are added to the final state, each will be assigned
unambiguously to a jet.
\item[(b)] \underline{The ``iterative-cone'' algorithm} \newline
Both CDF and D0 use this algorithm. 
While it is clearly based on the ``fixed-cone''
jet algorithm, there are important additions. The algorithm is given
by
\begin{itemize}
\item[1.] Form a cluster list, ordered by $E_T$.
\item[2.] Select the highest unassigned $E_T$-cluster, and draw a cone
          of radius R around the axis of this cluster. Calculate the
          transverse jet energy and a new jet-axis by merging the
          clusters in the cone as in eq.~\ref{eq:ETmerge}.
\item[3.] Draw a new cone around the new jet-axis. 
          Recalculate the jet-axis using the clusters 
          in the new cone. Repeat this step until a stable jet-axis 
           is found.
\item[4.] If there are clusters not yet assigned to at least one jet,
          go to step 1.
\item[5.] Check for overlapping clusters, 
          i.e. clusters assigned to two or more
          jets. If overlaps occur, one has to decide whether 
          to merge the jets or to assign the
          overlapping clusters to separate jets. 
          CDF and D0 have different methods
          for doing this. CDF merges the 
          jets if any of the overlapping jets shares more than 75\% of
          its $E_T$. Otherwise each shared cluster
          is assigned to the jet to whose axis it is
          closest in $\eta$-$\phi$ space.
          D0 merges the jets if any jet
          shares more that 50\% of its
          transverse energy. Otherwise the 
          shared transverse energy is divided equally
          between the two jets.  
\item[6.] Once all clusters have been uniquely assigned to jets, 
          the final jet parameters are calculated, but
          not using the $E_T$-weighted scheme of eq.~\ref{eq:ETmerge}.
          For both CDF and D0, the energy and momentum 3-vector are calculated
          by simply adding the 4-vectors of the clusters assigned to
          the jet, and the direction of the
          jet is given by the sum of the momentum 3-vectors.
          CDF computes the transverse energy
          of the jet as $E\sin\theta$, where $E$ is the energy
          calculated above, and $\theta$ is the polar angle of the jet
          direction. 
          D0 computes the transverse energy as the scalar sum of the
          transverse energies of the component clusters. 
          It is worth mentioning that
          ref. \cite{GK} recently argued that the 
          D0 procedure of defining the
          final jet parameters leads to large perturbative corrections
          and therefore should not be used.
\item[7.] Apply the appropriate minimum transverse energy and rapidity cuts 
          to the entries in the jet-list to find the final set of
          jets.
\end{itemize}

Note that in this case, unlike the ``fixed-cone'' algorithm, 
a lot of the physics is missing in the 3-parton final state, 
where there is never an iteration nor is there ever shared energy. 
To get all the basic physics one needs at least 
4 parton final states, or in other words NNLO 2-jet, 
NLO 3-jet or LO 4-jet production.
In fact for NLO 2-jet and LO 3 jet the ``iterative-cone''
algorithm is identical to the ``fixed-cone'' 
algorithm.
\item[(c)] \underline{The ``EKS'' algorithm} \newline
The fact that that the NLO 2-jet 
calculation does not contain all the needed
physics in the jet algorithms used by CDF and D0
inspired the authors of ref. \cite{EKSalg} to
introduce an ``improved'' algorithm which 
phenomenologically modeled the missing physics.
Because this is a theoretical algorithm we 
will describe it in terms of partons.
In NLO 2-jet production we have only to consider the 
3 parton final state. The algorithm
is then very simple:
\begin{itemize}
\item[1.] Consider the possible 2-parton configurations by 
          calculating their
          $E_T$-weighted jet axis as if they were clustered.
\item[2.] If both partons are within the 
          cone size $R$ of the hypothetical jet axis
          they are merged into a single jet.
\item[3.] Go to step 1 until all 2 parton configurations have 
          been considered.
\item[4.] Apply the appropriate minimum transverse energy and rapidity cuts 
          to the entries in the jet-list to find the final set of
          jets.
\end{itemize}
Note that this maximizes the energy in 
the cone and simulates the ``iterative-cone''
algorithm by assuming that it always find the 
optimum jet-axis to maximize the energy in a jet. 
This in fact overestimates the clustering effects of the 
``iterative-cone'' algorithm. To correct for this an additional 
parameter called $R_{sep}^{(2)}$ was introduced \cite{EKSRsep}.
With this parameter one can impose the additional constraint that
only 2-parton pairs separated by less than $R\times R_{sep}^{(2)}$ can be
clustered. Experimentally it was found that $R_{sep}^{(2)}=1.3$ 
worked best for $R=0.7$ \cite{CDFRsep}.
Note that the quantity $R_{sep}^{(2)}$ has no equivalent in
experimental jet algorithms and is a purely 
phenomenological quantity.
The $R_{sep}^{(2)}$ prescription was tuned to the NLO 2-jet calculation,
and there are many possible ways to extend it to the NLO 3-jet
calculation. We choose to do the following:
\begin{itemize}
\item[1.] Consider the possible 3-parton configurations by calculating their
          $E_T$-weighted jet axis as if they were clustered.
          If the three partons are within $R$ of the hypothetical
          jet axis and each pair of partons are separated by less than
          $R\times R_{sep}^{(3)}$
          they are merged into a single jet.
	  Repeat this step until all 3 parton 
          configurations have been considered.
\item[2.] Consider the possible 2-cluster configurations by calculating their
          $E_T$-weighted jet axis as if they were clustered.
          If both partons are within $R$ of the hypothetical jet axis
          and are separated from one another by less than $R\times
          R_{sep}^{(2)}$ they are merged into a single jet.
	  Repeat this step until all 2 cluster configurations 
          have been considered.
\end{itemize}
It is possible for two 2-parton clusters to overlap.  These situations
are resolved in the following fashion:
\begin{itemize}
\item[3.] If the shared parton contributes more than 75\% of the $E_T$
          of either jet, all three partons are merged.  If not, the
          shared parton is assigned to the jet to whose axis it is
          closest in $\eta$-$\phi$ space.
\item[4.] Apply the appropriate minimum transverse energy and rapidity cuts 
          to the entries in the jet-list to find the final set of
          jets.
\end{itemize}
Note that our implementation of the $R_{sep}$ parameters and overlap
resolution condition are {\it ad hoc\/}, not tuned to the data as
$R_{sep}^{(2)}$ was for the NLO 2-jet calculation.  For the NLO 3-jet
calculation, it could be that $R_{sep}^{(3)}$ should take on a
different value than $R_{sep}^{(2)}$, that a different overlap
resolution prescription will be preferred, or that additional
parameters will be needed to accurately describe the data.
\item[(d)] \underline{The ``$K_T$'' algorithm} \newline
This algorithm finds its roots in the $e^+e^-$ environment. 
Its adaptation to the $p\bar p$ environment was proposed 
in ref. \cite{KtalgTh}. The algorithm is currently
under study in CDF and D0 \cite{KtalgExp}. Our implementation
is based on ref. \cite{PbarPSeym}:
\begin{itemize}
\item[1.] For each cluster, $i$, define a ``closeness'' to the beam as
          $d_{ib} = E_{Ti}R_b$. For each pair of clusters $i$, $j$,
          define their closeness to one another as $d_{ij} =
          \min\{E_{Ti}, E_{Tj}\}\Delta R_{ij}$.
\item[2.] Choose the cluster closest to the beam ($\min\{d_{ib}\}$).
          If $\min\{d_{ij}\} < d_{ib}$, merge $j$ into $i$, 
          and remove $j$ from the cluster list.  If all $d_{ij} >
          d_{ib}$, jet $i$ is said to be ``complete.''
\item[3.] Go to step 1 until all jets are complete.
\item[4.] Apply the appropriate rapidity and transverse energy cuts to
          select the final set of jets.
\end{itemize}
All of the basic physics involved in the $K_T$ clustering algorithm
was already present in the 3 parton final states.  Like the fixed cone
algorithm, the $K_T$ algorithm unambiguously assigns additional
partons to jets, no matter how many are added.
\end{itemize}

The numerical stability of the four jet algorithms 
is related to the degree in which
the algorithm is sensitive to soft radiation, 
or in other words the infrared 
stability of the particular algorithm. 
For the method of resolved partons,
as is used in this paper, 
infrared stability is related to the extent to which the results are
independent of the the resolution parameter $s_{min}$. 
This dependence is shown in fig.~\ref{fig:smin} and will be discussed
in the next section. 

\section{Numerical Results}
The calculation presented in this paper includes only 
the $gg\rightarrow ggg$ and $gg\rightarrow gggg$ contribution to the
NLO 3-jet cross section. This means that any comparison with
experimental results would be premature. However, there are several
issues we can address in the context of investigating the numerical
applicability of the resolved parton approach.  First, we can get a
first impression of the size of the radiative corrections in the
inclusive 3-jet cross section by comparing the all-gluon LO 3-jet
results with the NLO 3-jet results.  Second, we can start to look at
questions related to the jet-algorithms and to what extent observables
depend on the choice of algorithm. We will look at two particular sets
of observables. The first set is the transverse energy distribution of
the leading, second and third jet in the event. The second set
involves the transverse energy fraction of the leading, second and
third jet.

For all numerical results in this section we used the CTEQ3M
\cite{CTEQ3M} parton distribution functions (PDF's), a fixed
renormalization/factorization scale of 100 GeV 
and a center of mass energy of the $p\bar{p}$-system equal to 1800
GeV. The fixed scale is needed at NLO because we calculate the
gluons-only cross section. The full PDF's, including the quarks, are
evolved up to $Q=100$ GeV. The input gluon PDF is then taken at this
scale and not evolved any further by fixing the factorization scale at
100 GeV. In this manner we get a consistent cross section with only
gluons (i.e. taking the number of flavors equal to zero) at NLO.  To
select events we required at least one jet with $E_T > 50$ GeV in the
rapidity region, $|\eta| < 4$.  Additional jets were required to have
$E_T > 20$ GeV and rapidity in the range $|\eta| < 4$.  Only events
with at least three jets in the final state were selected. The cone
sizes were chosen differently per algorithm such that they give
approximately the same cross section.  The ``iterative-cone''
algorithm uses the same cone size of 0.7 as is usually chosen
experimentally.  In the ``EKS'' algorithm the cone size was chosen to
be 0.7 with $R_{sep}^{(2)}=R_{sep}^{(3)}=1.3$ as is common in the NLO
2-jet calculations. In order to accommodate the larger ``effective''
cone of the two previous algorithm we chose the ``fixed-cone'
algorithm to have a larger cone, $R=0.7\times 1.3 = 0.91$.  Finally
for the ``$K_T$''-clustering algorithm the closeness parameter is set
to $R_b=1.0$ (note that this quantity is not really a cone size).

\begin{figure}[pt]
\hbox{\epsfxsize=225pt\epsfbox{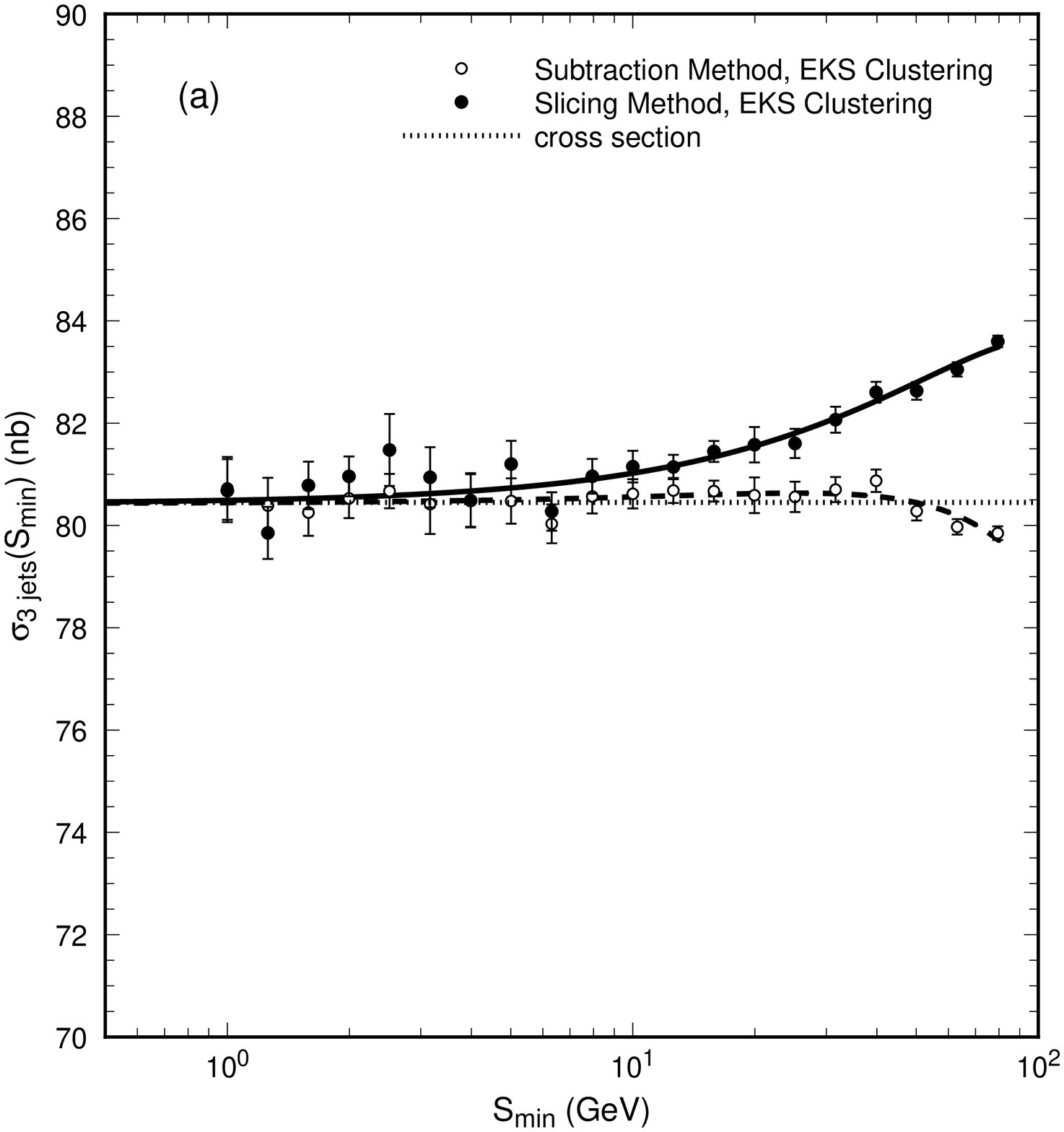}
      \epsfxsize=225pt\epsfbox{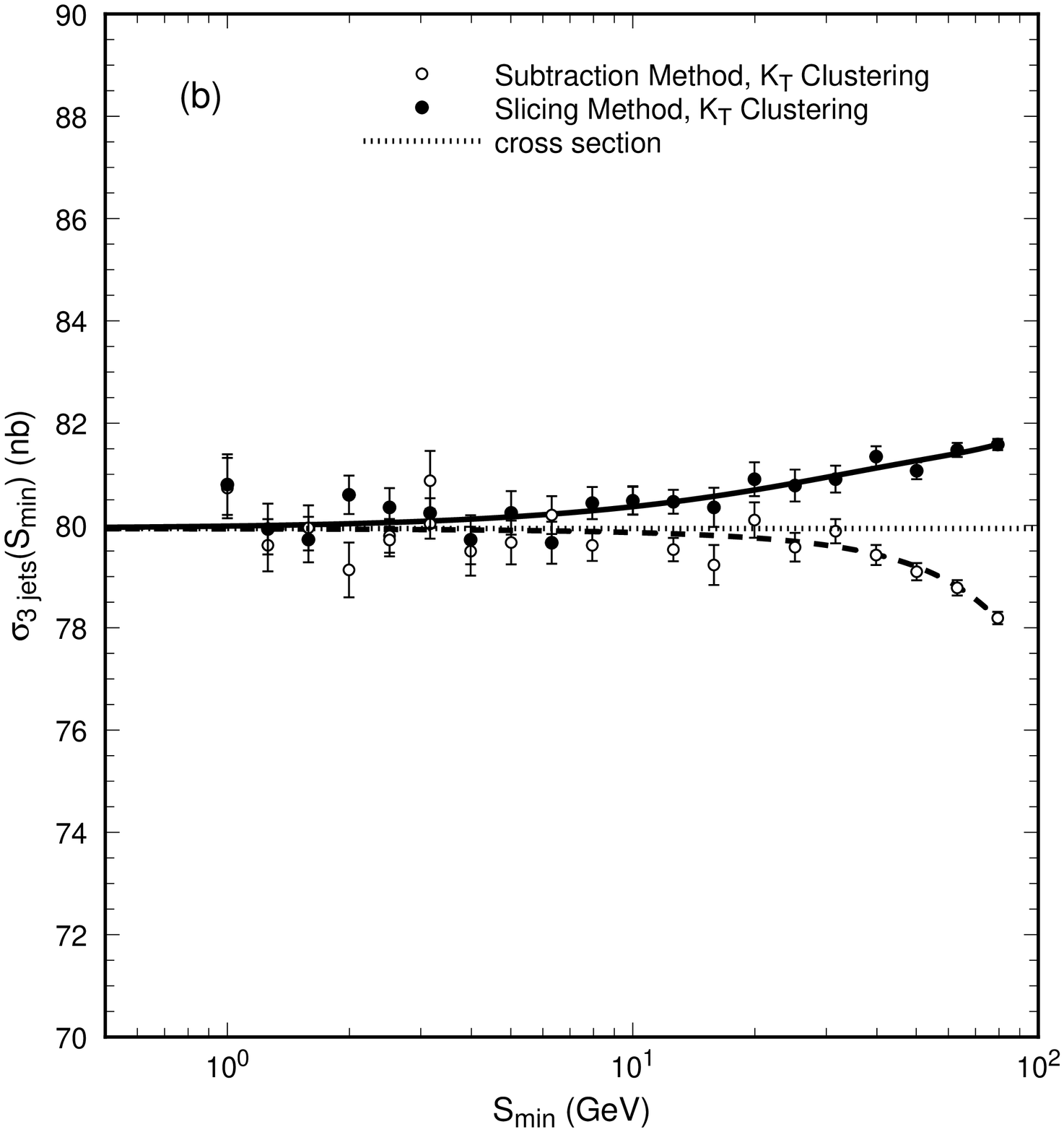}}
\vspace{25pt}
\hbox{\epsfxsize=225pt\epsfbox{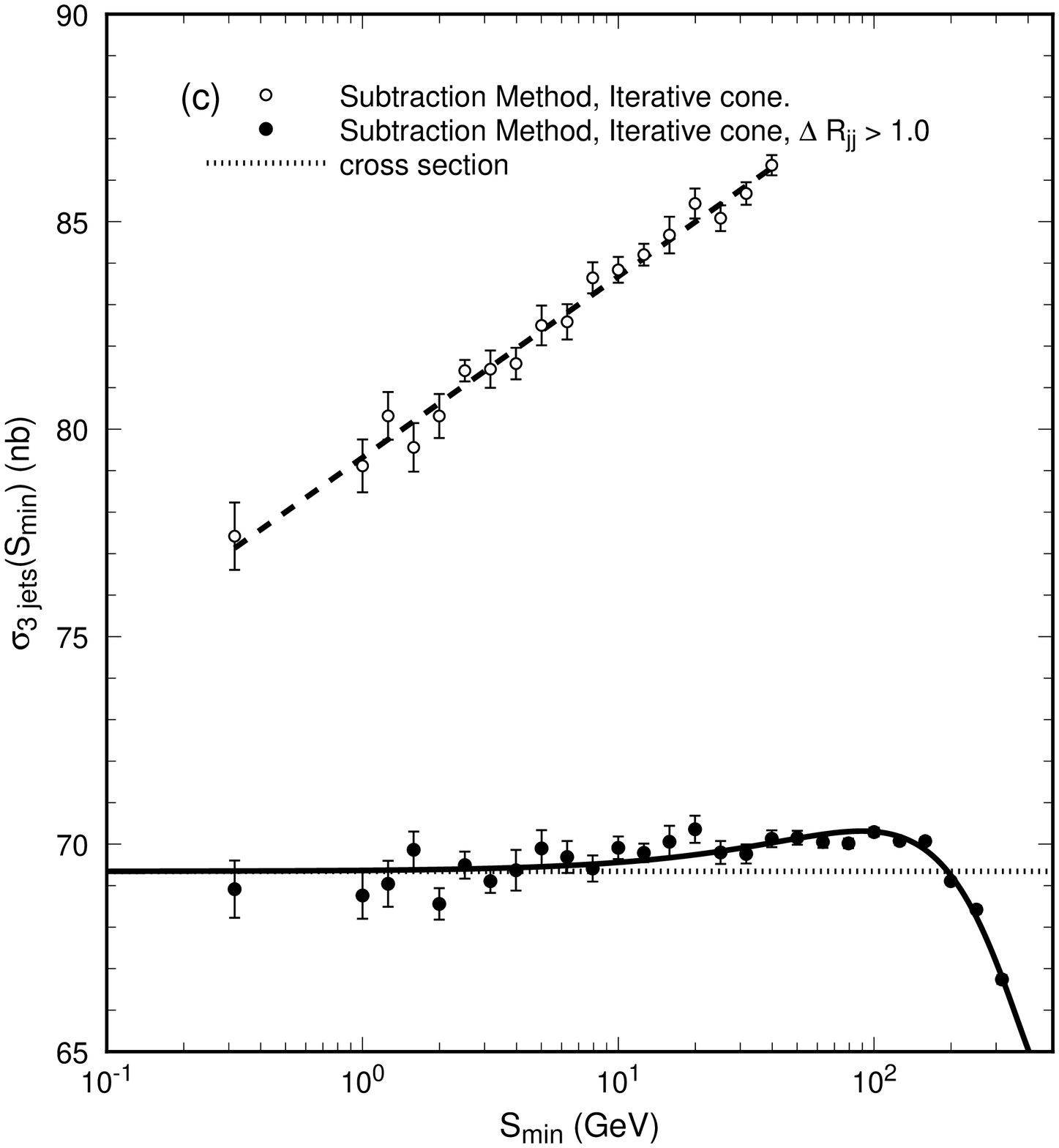}
      \epsfxsize=225pt\epsfbox{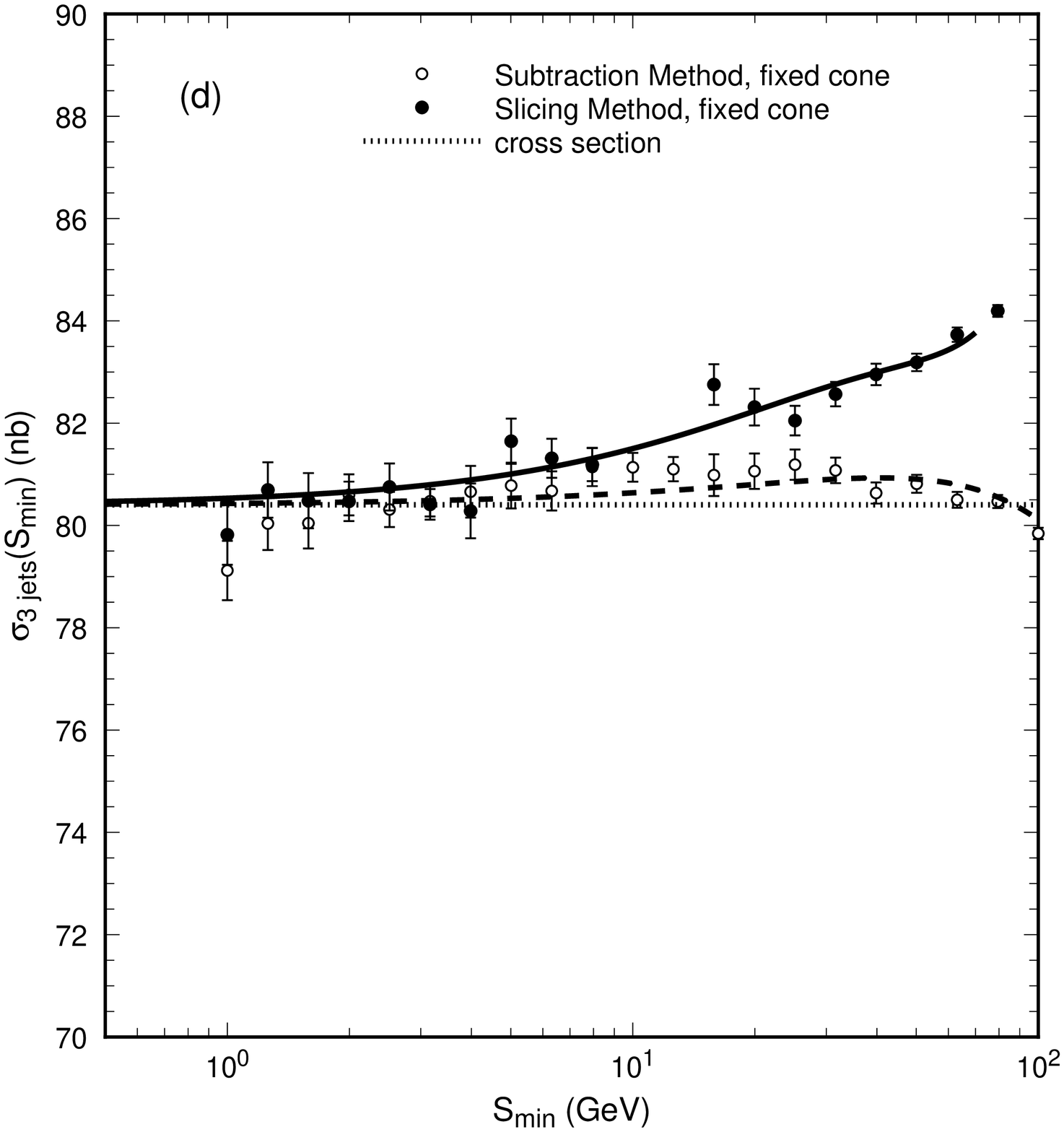}}
\caption[]
{The $s_{min}$-dependence of the cross section for the different
 jet algorithms and numerical methods.}
\label{fig:smin}
\end{figure}

The first issue to be considered is the $s_{min}$-dependence of
the cross section and the determination of the range in which we can
choose its value such that the approximations made in the different
numerical methods are valid.  The results are shown in
fig.~\ref{fig:smin} for both
the slicing and subtraction method (the exact method has not yet been
implemented) and all four types of jet algorithms.  The first thing to
notice is that the behavior of the iterative cone algorithm is
quantitatively different from that the three other algorithms. The
other three algorithms behave as expected and it is clear how to choose
$s_{min}$ for them.  For the slicing method one has to choose
$s_{min}$ smaller than 1 GeV$^2$ in order to get the correct
answer. As expected the subtraction method allows us to choose larger
values of $s_{min}$, though the value should still not be larger than
10 GeV$^2$. For the results presented later in this section we will
use the subtraction method with $s_{min} = 2.5$ GeV$^2$. 

We now consider the iterative cone algorithm. As can be seen in
fig.~\ref{fig:smin}c, the cross section does not become independent
from the resolution parameter, even at very small values of $s_{min}$.
In fact the behavior fits very well to a logarithmic dependence on the
resolution parameter.  This means that the algorithm is not infrared
safe in that we can change the jet multiplicity by adding a soft
parton somewhere in the event.  It is obvious that this can occur when
we can have three parton configurations in which two of the partons
are slightly more than the cone size $R$ apart balancing the leading
third parton.  For the tree level and virtual contributions this is a
three jet event.  The situation should not change if we add a soft
parton in between the two nearby partons, and in fact it does not
change for any of the jet algorithms besides the iterative cone. The
soft parton gets clustered with one of the hard partons, slightly
changing the jet parameters, but not affecting the jet multiplicity.
In the case of the iterative cone however, one of the two hard partons
will cluster with the soft parton thereby shifting its jet-axis to
within R from the other parton. Because of the iterative nature of
this algorithm the two clusters will subsequently be merged further
into a single jet yielding a two jet final state.  Thus, we have
changed the jet multiplicity by adding an arbitrarily soft parton to
the event. As a result the algorithm is infrared unstable and cannot
be used within the context of perturbative QCD. Experimentally this
means that the jet algorithm depends on the implicit soft cut-offs in
the detector, e.g. granularity of the detector, cluster cut-off and
ultimately hadron masses.  In other words, the jet multiplicity
depends on the ability of the detector to resolve and measure soft
hadrons.  It is clear that we cannot use this algorithm within the NLO
calculation.  Note that this result does not make the one- and two-jet
inclusive cross section infrared unstable since in those cases we do
not have to resolve three-jet configurations.  Both CDF and D0 have
compared their multi-jet data (i.e. more than two jets in the final
state) with LO monte carlo's \cite{D0alg,CDF2}. It is interesting to
note that the experiments have in fact added an additional cut to
their multi-jet cross section in order to make these comparisons. This
cut requires all the jets in the event to be further apart then their
cone-size of $R=0.7$. For CDF this cut was $\Delta R_{jj} > 1.0$,
while for D0 the requirement is $\Delta R_{jj} > 1.4$. This additional
requirement to the jet algorithm changes the $s_{min}$-dependence of
the cross section dramatically, as can be seen clearly in
fig.~\ref{fig:smin}c. In fact the behavior is now very similar to the
other three algorithms. This is no surprise since with this additional
selection cut the infrared instability is removed. This means that the
iterative cone algorithm needs to be augmented with a jet separation
cut in order to be an infrared safe jet algorithm.

\begin{figure}[pt]
\hbox{\epsfxsize=225pt\epsfbox{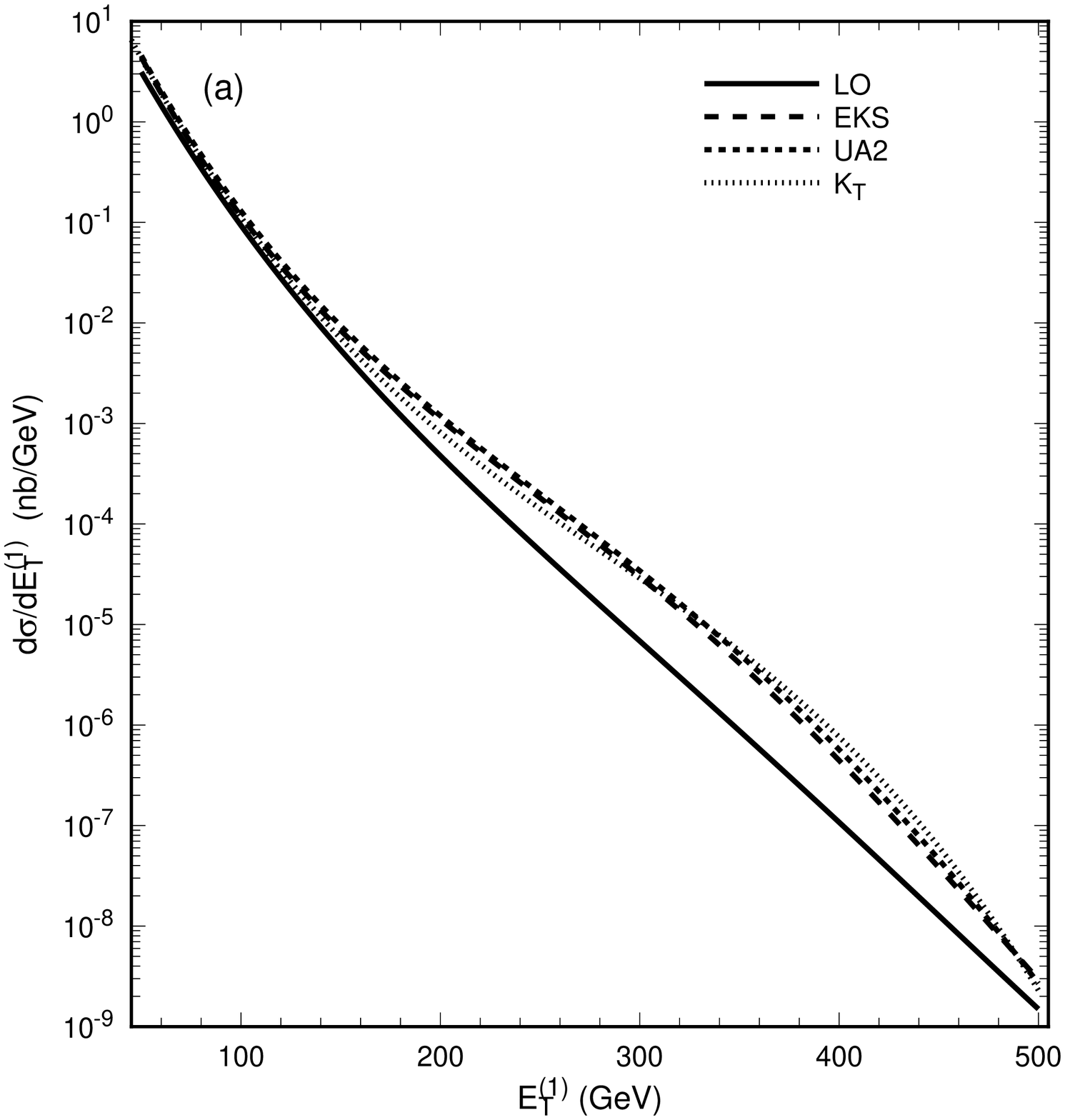}
      \epsfxsize=225pt\epsfbox{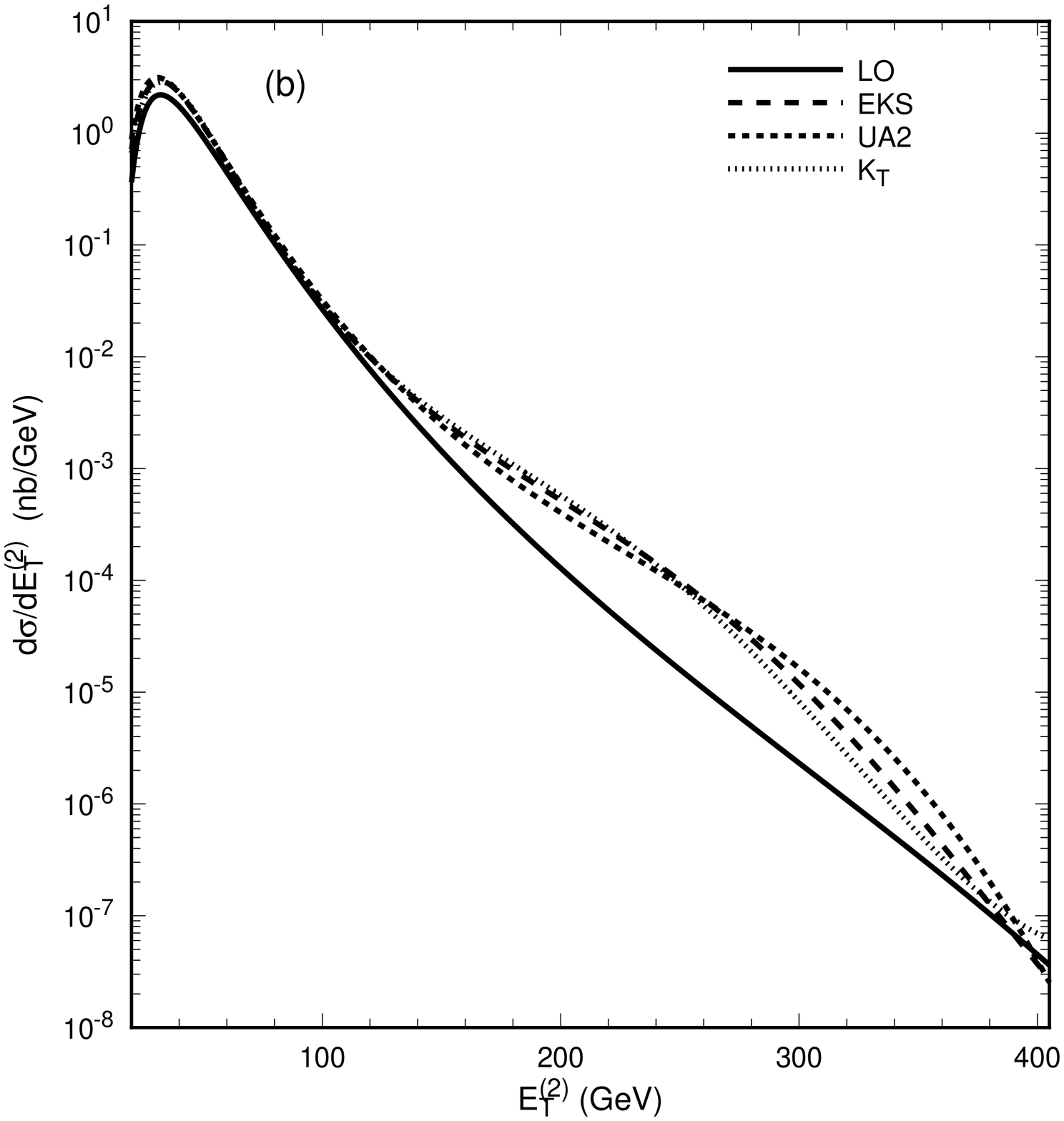}}
\vspace{25pt}
\hbox{\epsfxsize=225pt\epsfbox{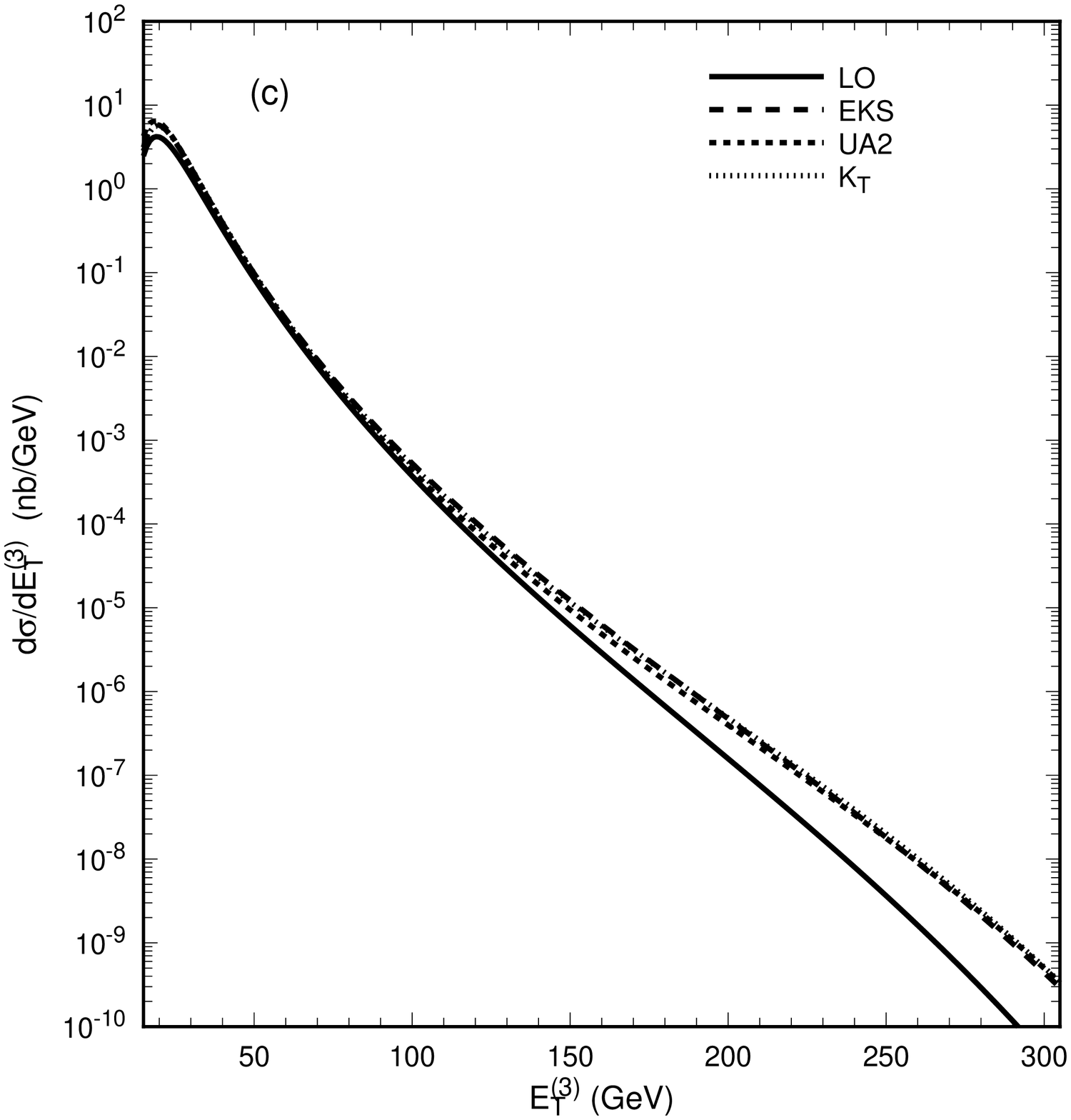}
      \epsfxsize=225pt\epsfbox{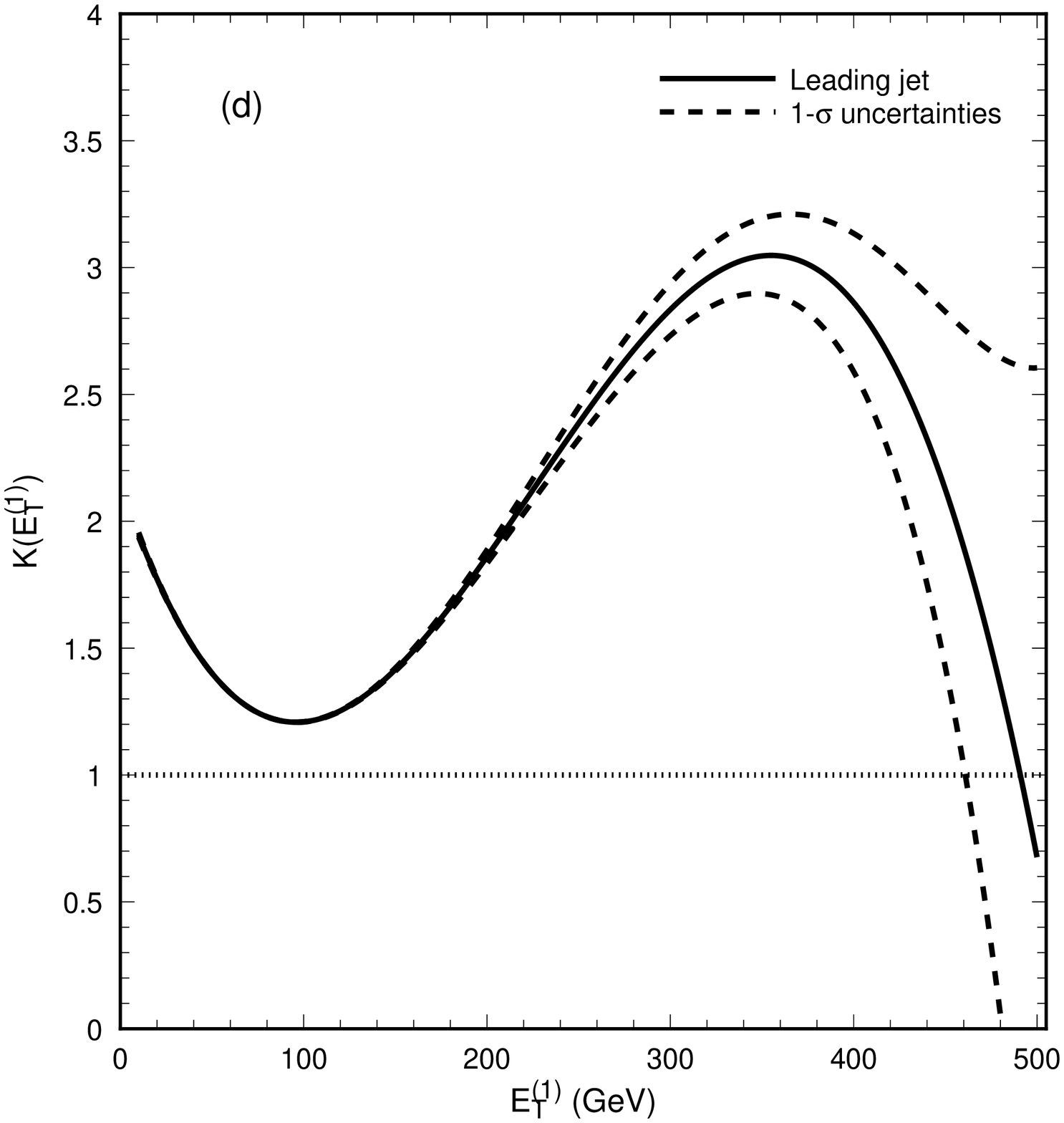}}
\caption[]
{The $E_T$-spectra of the (a) leading, (b) second and the (c) third jet.
 Fig. d contains the $K$-factor of the leading jet for the EKS clustering
 scheme.}
\label{fig:ETdist}
\end{figure}
The most basic distributions we can look at are the $E_T$-ordered
transverse energy distributions. These distributions are given in
fig.~\ref{fig:ETdist} for various jet algorithms. The curves are fits
to M.C. output and have a fit-uncertainty associated with them. The
fit uncertainty for the leading jet is shown in fig.~\ref{fig:ETdist}d
where the leading jet $K$-factor (i.e. the ratio of NLO over LO) is
given together with the 1-$\sigma$ boundary on the fit. The
uncertainties on the second and third jet are very similar in size and
$E_T$-dependence. As can be seen from figs~\ref{fig:ETdist}a,
\ref{fig:ETdist}b and \ref{fig:ETdist}c the differences between the
jet algorithms are small and stable, especially when taking the
fit-uncertainties into account.  The LO normalization is highly
uncertain because it is an $\alpha_S^3$-process and therefore very
dependent on the value of $\alpha_S$ (i.e. at LO the renormalization
scale choice).  The radiative corrections, however, show more
structure than a simple normalization shift. The radiative effects can
be quite substantial, with a $K$-factor as large as three for
$E_T=350$ GeV. There are two reasons for these large corrections. Note
that the minimum in the $K$-factor for leading jet occurs at $E_T=100$
GeV, exactly at the renormalization/factorization scale choice. This
is no accident.  Usually one would choose this scale to be
equal/proportional to the leading jet $E_T$.  For the gluons-only
process, however, this would require evolving the PDF's with
$n_f=0$. So, part of the 
large corrections away from $E_T=100$ GeV are due to the choice of
renormalization/factorization scale which generates large logarithmic
corrections at higher orders. The second reason is that we look at
gluons only, while evolving the PDF's to a scale of 100 GeV using both
quarks and gluons. This means the gluon content of the proton and
therefore the size of the radiative corrections in the gluons-only
case depend on the mass factorization scheme used in the PDF and
matrix element. Any conclusions on the radiative corrections in the
full case (i.e. including the quark processes) is therefore
premature. Note that in the $\overline{\mbox{MS}}$-scheme used in this
calculation the contribution of gluon initiated scattering at
$E_T=350$ GeV is very small. The scattering at such large momentum
transfers is dominated by $t$-channel quark scattering, making the
size of the gluons-only $K$-factor irrelevant.

\begin{figure}[pt]
\hbox{\epsfxsize=225pt\epsfbox{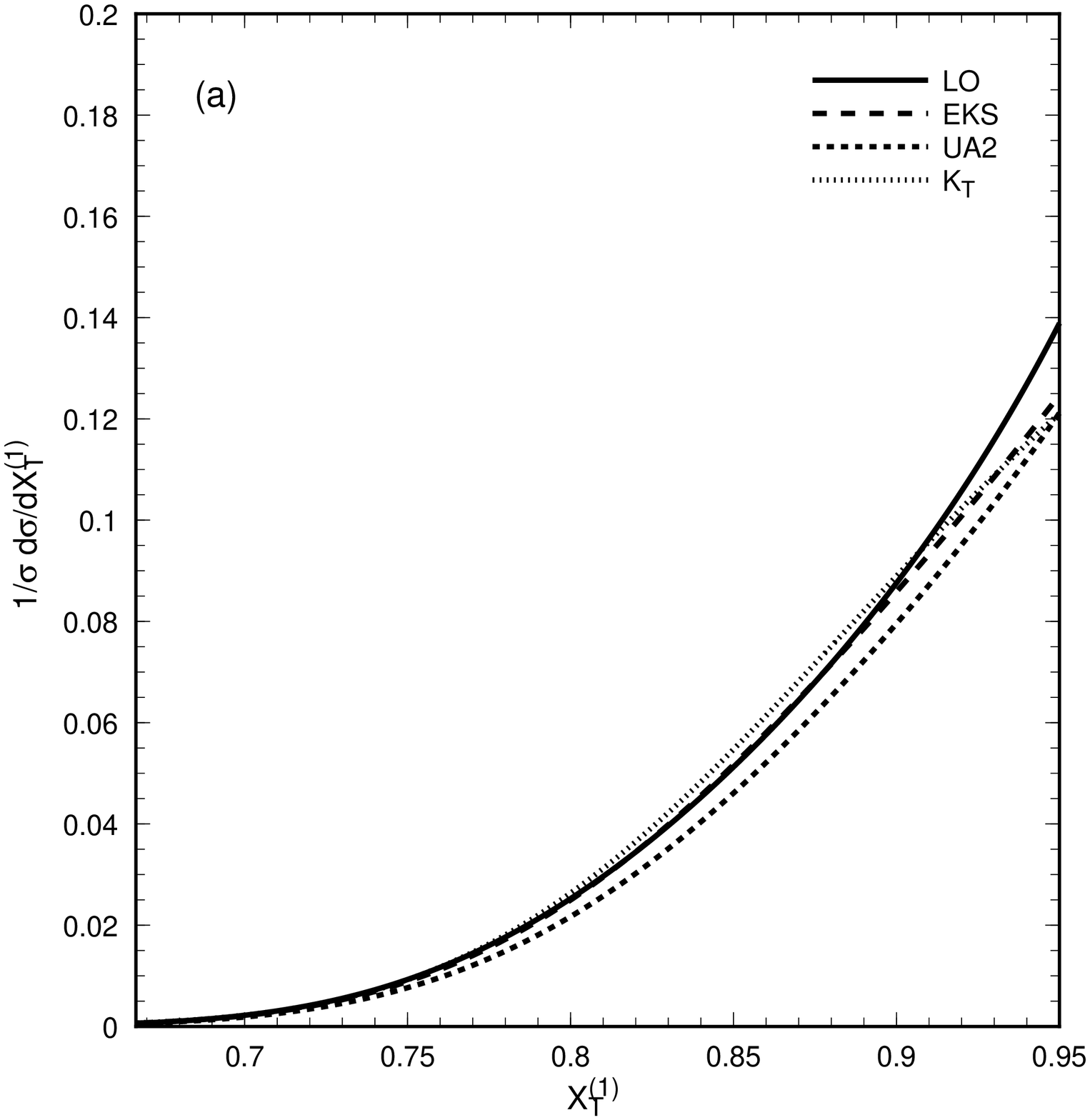}
      \epsfxsize=225pt\epsfbox{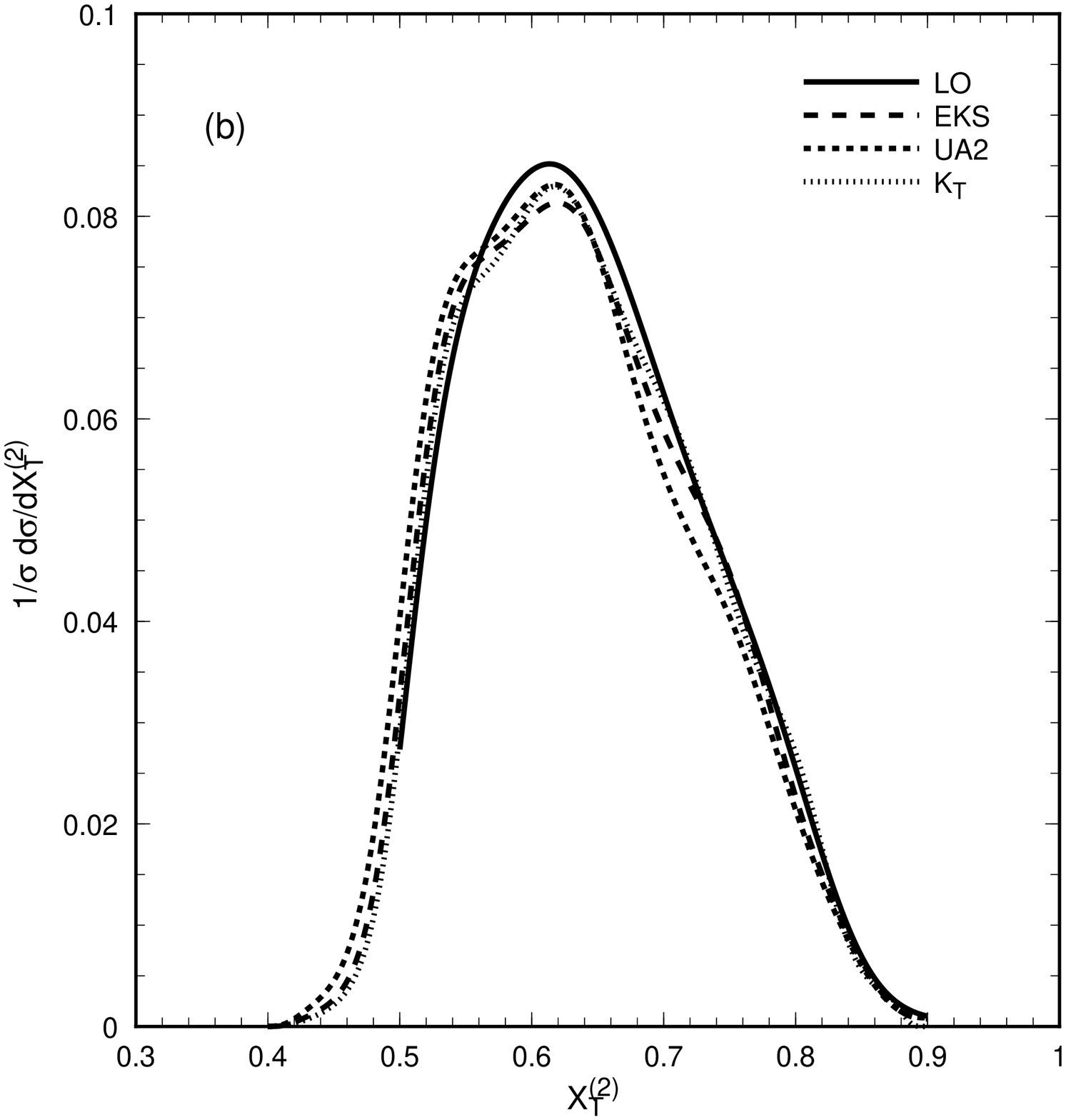}}
\vspace{25pt}
\hbox{\epsfxsize=225pt\epsfbox{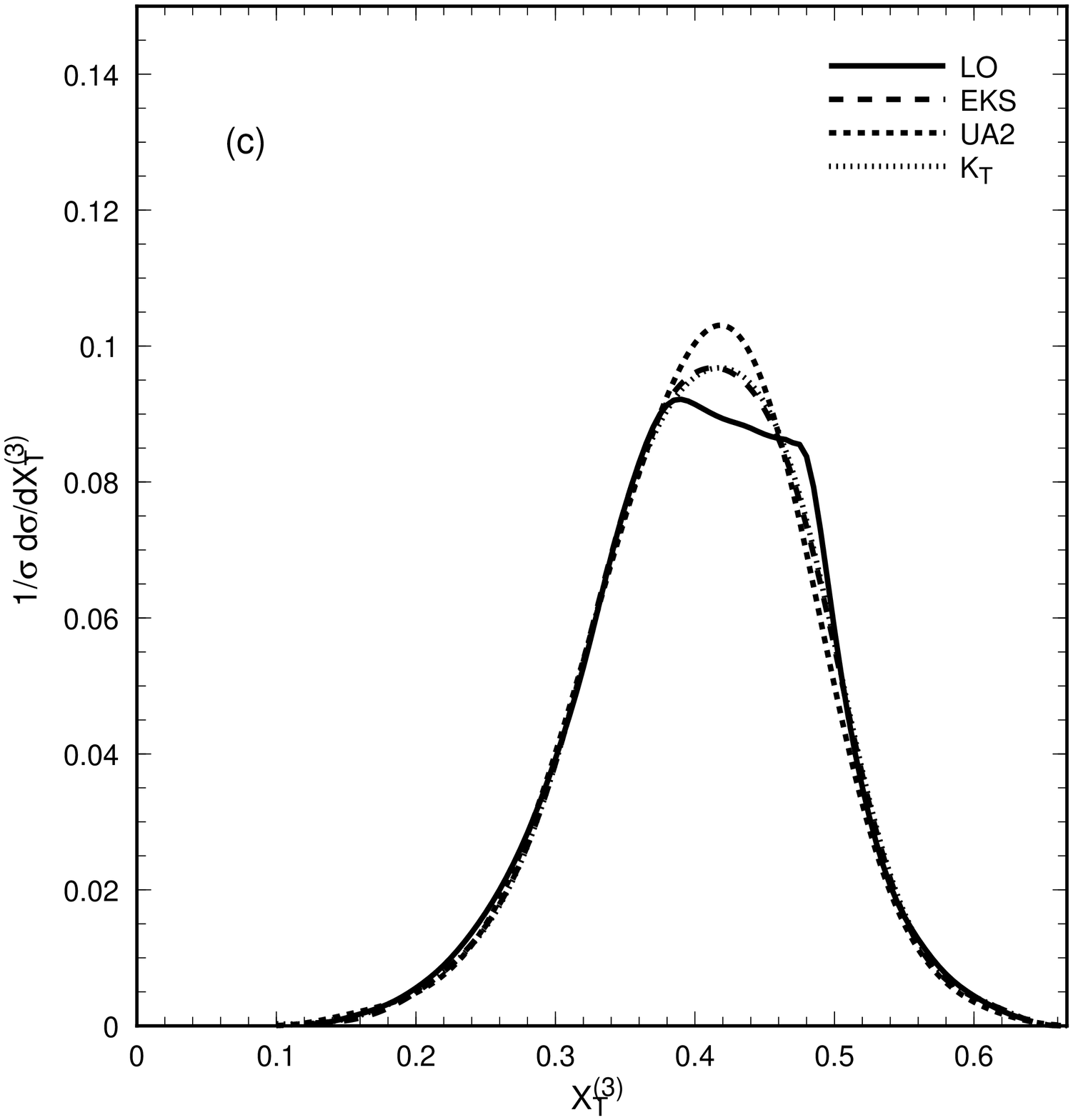}
      \epsfxsize=225pt\epsfbox{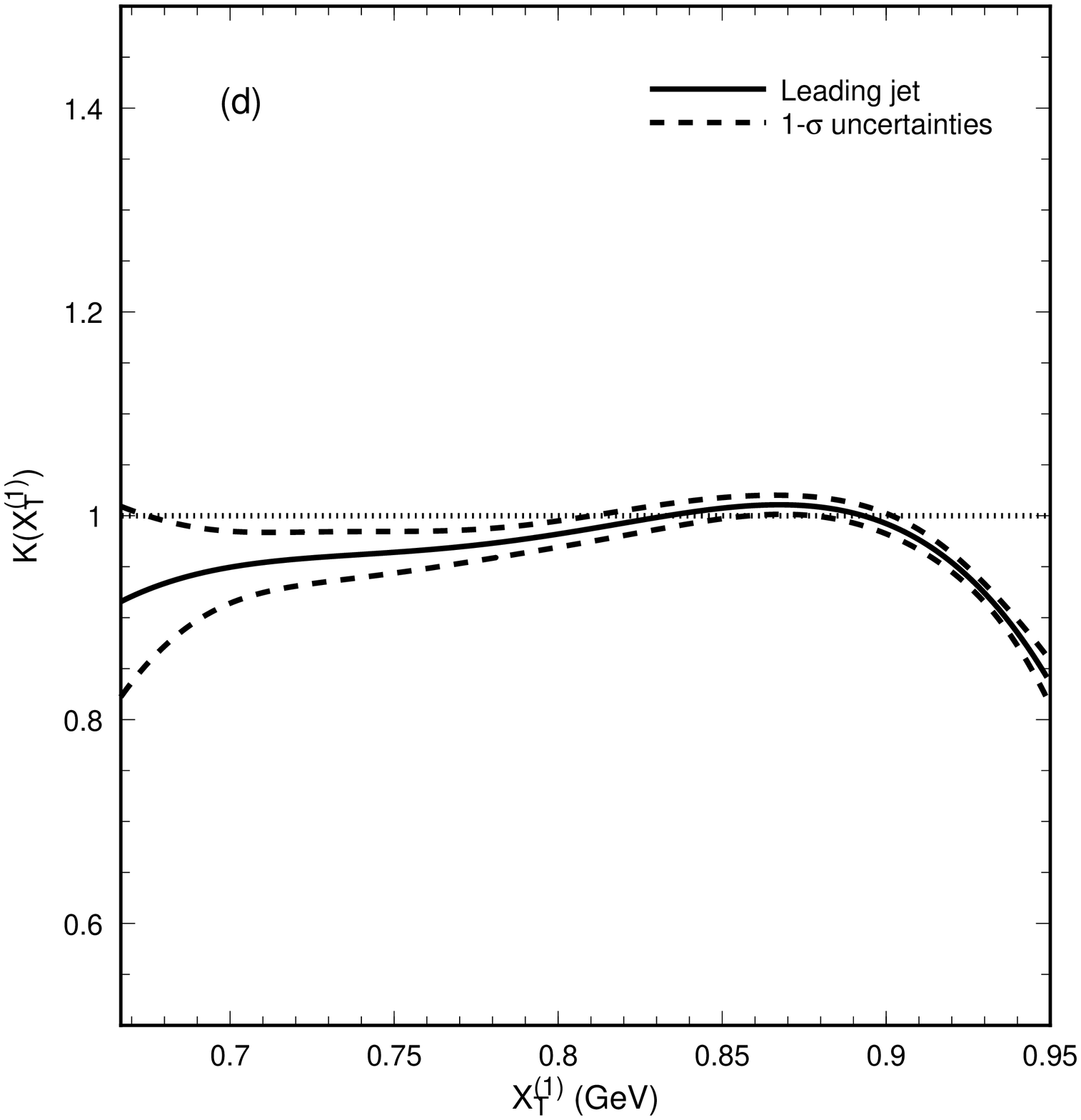}}
\caption[]
{The $X_T$-spectra of the (a) leading, (b) second and the (c) third jet
 and (d) the $K$-factor for the leading jet as a function of $X_T$.}
\label{fig:ETfrac}
\end{figure}

The final observable we will look at in our investigation of the
stability of the NLO 3-jet event generator is the transverse energy
fraction $X_T^{(i)}=2 E_T^{(i)}/\sum_{j=1}^3 E_T^{(j)}$ of the three
leading jets (in transverse energy) in the event.  These are different
from the usual observables used by the experimentalists (see e.g. the
CDF papers \cite{CDFalg,CDF2}). They look at the energy fraction
$X^{(i)}=2 E^{(i)}/M_{jjj}$ where the energies are defined in the
center of mass frame of the collision and $M_{jjj}$ is the invariant
mass of the three leading jets.  We have chosen the transverse energy
fractions because they do not require the determination of the
center of mass reference frame. At NLO, this frame is strongly
dependent on the ability to detect forward radiation, making the NLO
prediction rather unstable and detector dependent. The transverse
energy fraction, on the other hand, behaves more stably and radiative
effects are small. This can be seen in fig.~\ref{fig:ETfrac} where the
normalized LO and NLO transverse energy fraction distributions are
plotted for several jet algorithms. Also shown is the $K$-factor for
the normalized $X_T^{(1)}$-distribution together with its
fit-uncertainties. The radiative corrections for these distributions
are in general small, except at the edge of LO phase space where the
jet-algorithm sensitivity also becomes large. (At LO the transverse
energy fractions are constrained to $2/3 < X_T^{(1)} < 1$, $ 1/2 <
X_T^{(2)} < 1$ and $0 < X_T^{(3)} < 2/3$, not taking any $E_T$-cuts
into account.)  The NLO 3-jet event generator is capable of predicting
these distributions accurately enough for comparisons with
experiments.

\section{Conclusions}
In this paper we have presented results on the purely gluonic
contribution to the NLO 3-jet cross section.  All of the techniques
used can be readily applied to the quark contributions. Several
techniques to isolate the soft/collinear contributions were explored
and their numerical effects investigated.

All of the relevant experimental jet algorithms were implemented in
the NLO 3-jet event generator and their radiative effects studied. For
the iterative cone algorithm it was necessary to augment the algorithm
with an additional jet separation cut in order to obtain infrared
stability. Both CDF and D0 already apply such a cut in their multijet
analysis, though the reason is the inefficiency of the
cluster algorithm instead of the theoretically motivated removal of
the infrared instability.  The other jet algorithms behaved properly
and no additional cuts were needed.

The NLO 3-jet event generator was applied to several distributions 
and it was demonstrated that one could obtain useful results
which can be compared to the experimental data, once the quark matrix
elements are included.

\section*{Acknowledgments} We would like to thank the Fermilab
lattice QCD group for the kind use of the ACPMAPS supercomputer, on
which the calculations in this paper were performed, and George
Hockney for his assistance in using it.  Fermilab is
operated by Universities Research Association, Inc., under contract
DE-AC02-76CH03000 with the U.S. Department of Energy.


\end{document}